\documentclass[11pt]{article}
\usepackage{amssymb}

\begin{document}

\title{ Effects of the R-parity violation in the minimal supersymmetric standard model
on dilepton pair production at the CERN LHC \footnote{Supported by National Natural
Science Foundation of China.}} \vspace{3mm}

\author{{ Yin Jun$^{2}$, Ma Wen-Gan$^{1,2}$, Wan Lang-Hui$^{2}$, and Zhang Ren-You$^{2}$}\\
{\small $^{1}$ CCAST (World Laboratory), P.O.Box 8730, Beijing
100080, P.R.China}\\
{\small $^{2}$ Department of Modern Physics, University of Science and Technology}\\
{\small of China (USTC), Hefei, Anhui 230027, P.R.China}}

\date{}
\maketitle \vskip 12mm

\begin{abstract}
  We investigate in detail the effects of the R-parity lepton number
violation in the minimal supersymmetric standard model (MSSM) on
the parent process $pp \rightarrow  e^+ e^- + X$ at the CERN Large
Hadron Collider (LHC). The numerical comparisons between the
contributions of the R-parity violating effects to the parent
process via the Drell-Yan subprocess and the gluon-gluon fusion
are made. We find that the R-violating effects on $e^+ e^-$ pair
production at the LHC could be significant. The results show that
the cross section of the $ e^+ e^-$ pair productions via
gluon-gluon collision at the LHC can be of the order of $10^2$ fb,
and this subprocess maybe competitive with the production
mechanism via the Drell-Yan subprocess. We give also
quantitatively the analysis of the effects from both the mass of
sneutrino and coupling strength of the R-parity violating
interactions.
\end{abstract}

\vskip 5cm


\vfill \eject

\baselineskip=0.36in

\renewcommand{\theequation}{\arabic{section}.\arabic{equation}}
\renewcommand{\thesection}{\Roman{section}}
\newcommand{\nb}{\nonumber}

\makeatletter      
\@addtoreset{equation}{section}
\makeatother       

\section{Introduction}
The extensions of the standard model (SM) have been intensively
studied over the past years\cite{beyondSM}. The minimal
supersymmetric standard model (MSSM) is one of the most attractive
ones among the general extended models of the SM. In the usual
MSSM the R-parity is conserved. Here the R-parity is defined as
$$
R_{p}=(-1)^{3B+L+2S}.
$$
We can see that the lepton-number or baryon-number violating
interactions can induce R-parity violation (${\rlap/{R}}_p$) . In
the usual supersymmetric (SUSY) extension models, R-parity
conservation \cite{rdef} is imposed due to two reasons. One is to
retain the electroweak gauge invariance of the SM. the other is to
solve the proton decay problem, since the R-parity violation leads
to unacceptable short proton lifetime. But the most general SUSY
extension of the SM should contain R-parity violating
interactions. Until now have we been lacking in credible
theoretical argument and experimental tests for $R_p$
conservation, so we can say that the $R_p$ violation would be
equally well motivated in the supersymmetric extension of the SM.
Up to now we have experimentally only some upper limits on
${\rlap/{R}}_p$ parameters. Therefore, it is necessary to continue
the work on finding ${\rlap/{R}}_p$ signal or getting further
stringent constraints on the ${\rlap/{R}}_p$ parameters in future
experiments.
\par
Recent experimental and theoretical works have demonstrated that
the dilepton production processes in hadron collisions ($p\bar{p}
\rightarrow l\bar{l}+X$ and $pp \rightarrow l\bar{l}+X$
$(l=e,\mu)$), are very important \cite {U.B}, since there is a
continuous c.m.s energy distribution of the colliding partons
inside protons(and anti-protons) at hadron colliders. These
dilepton production channels at hadron colliders can be used to
study the parton distribution functions(PDFs) at small x values
\cite{F.A1} \cite{F.A2}, to determine the $W$ mass \cite{F.A3}
\cite{S.A} \cite{R.B}, and to extract the effective weak mixing
angle\cite{C.A} and the information on the width of the $W$ boson
\cite{S.A2} \cite{F.A4},...etc. The most attractive purpose is
that at the c.m.s hadron colliding energy region beyond the
$m_{Z^0}$, one can probe the new physics beyond the SM, such as
large extra dimensions \cite{F.A5} \cite{F.A7}, extra neutral
gauge bosons\cite{F.A6}, R-parity violation\cite{chou} and
composite quarks and leptons\cite{F.A8}. Therefore, we can
conclude that at the upgrade Fermilab Tevatron Run II with
integrate luminosity 2-20$fb^{-1}/year$ at $\sqrt{s}=2~TeV$ and
the CERN Large Hadron Collider (LHC) with 100$fb^{-1}/year$ at
$\sqrt{s}=14~TeV$, it would be possible to extract the new physics
effects beyond the SM by investigating the dilepton production
processes.
\par
The dilepton pairs can be produced via the Drell-Yan subprocess
and gluon fusion at hadron colliders. The analysis of the
sensitivity to ${\rlap/{R}}_p$ couplings in measurement on the
Drell-Yan production processes $ pp \rightarrow u_j u_j(d_k d_k)
\rightarrow l^+_i l^-_i+X$ at the LHC was studied in
Ref.\cite{chou}. But we suppose that the process $pp \rightarrow
gg \rightarrow l^+_i l^-_i+X$ could be important too due to the
large gluon luminosity in hadron colliders, although the dilepton
pair production via gluon fusion is an one-loop process.
\par
If we set the values of the R-parity violating parameters
concerned to be near the corresponding upper limits\cite{B.A}, we
can see that the major R-parity effects on cross section of
subprocess $gg \rightarrow l^+_{i}l^-_{i}(l_{i=1,2}=e,\mu)$ come
from the loops involving the third generation quarks. Then we
would have the conclusion that the cross section of $gg
\rightarrow l^+_{\alpha}l^-_{\alpha}$ is  approximately
proportional to
$\sum_{i=1}^{2,3}(\lambda'_{i33}\lambda_{i\alpha\alpha})^2$
($i,\alpha$ are the generation indices of the sneutrino and
dilepton, respectively). From the experimental constraints of
${\rlap/{R}}_p$ parameters\cite{B.A}, we have the upper
limitations on the related R-parity violating parameters:
$\lambda'_{133}\leq 1.4\times 10^{-3}\sqrt{m_{\tilde{b}}/100GeV}$,
$\lambda'_{333}\leq 0.45$, $\lambda'_{233}\leq 0.15 \times
\sqrt{m_{\tilde{b}}/100GeV}$, $\lambda_{311}\leq 0.062\times
\frac{m_{\tilde{e}_R}}{100GeV}$, and $\lambda_{211}\leq
0.049\times \frac{m_{\tilde{e}_R}}{100GeV}$. Due to
$\lambda_{ijk}=-\lambda_{jik}$, we have
$\lambda_{ijk}=0,~(for~i=j)$. Since the upper limitation of
$\lambda'_{133}$ is much smaller than those of $\lambda'_{233}$
and $\lambda'_{333}$, the cross section of $gg \rightarrow
\mu^+\mu^-$ is approximately proportional to
$(\lambda'_{333}\lambda_{322})^2$. And the cross section of $gg
\rightarrow  e^+ e^-$ is approximately proportional to
$(\lambda'_{333}\lambda_{311}+\lambda'_{233}\lambda_{211})^2$.
Then we can estimate that the cross section of $gg\rightarrow
 e^+ e^-$ may be several times larger than that of $gg \rightarrow
\mu^+\mu^-$, if the related R-parity violating parameters have the
values near the corresponding upper limits.
\par
In this paper we concentrated on finding the effects of R-parity
lepton number violating MSSM in the $ e^+ e^-$ pair production
processes via both Drell-Yan and gluon fusion subprocesses at the
CERN LHC. In section II we present the relevant model and the
calculation of the processes $pp \rightarrow e^{+}e^{-}+X$.
Analytical calculation is presented in section II. Numerical
results and discussion are given in section III, where the cross
sections in the minimal standard supersymmetric models with and
without R-parity violation will be compared. In section IV we give
a short summary.

\section{Relevant Theory and Calculations}
\par
In this section we review briefly the MSSM with R-parity lepton
number violation. All renormalizable supersymmetric $\rlap/R$
interactions can be introduced in the superpotential. The general
form of the super potential can be written as \cite{Yin}.
\begin{equation}
{\cal W} = {\cal W}_{MSSM}+{\cal W}_{\rlap/{R}}
\end{equation}
where ${\cal W}_{MSSM}$ represents the R-parity conserved term,
which can be written as
\begin{equation}
{\cal W}_{MSSM} = \mu \epsilon_{ij} H_i^1 H_j^2+\epsilon_{ij} l_I H_i^1 \tilde{L}_j^I
\tilde{R}^I-u_I (H_1^2 C^{JI \ast} \tilde{Q}_2^J-
     H_2^2 \tilde{Q}_1^J)\tilde{U}^I-d_I (H_1^1 \tilde{Q}_2^I -
     H_2^1 C^{IJ} \tilde{Q}_1^J) \tilde{D}^I
\end{equation}
and ${\cal W}_{\rlap/{R}}$ represents the term of R-parity
violation,
\begin{equation}
W_{\rlap/{R}} = \epsilon_{ij} (\lambda_{IJK} \tilde{L}_i^I \tilde{L}_j^J
     \tilde{R}^K+\lambda_{IJK}' \tilde{L}_i^I \tilde{Q}_j^J \tilde{D}^K+
     \epsilon_I H_i^2 \tilde{L}_j^I)+\lambda_{IJK}''\tilde{U}^I
     \tilde{D}^J \tilde{D}^K
\end{equation}
The soft breaking terms can be expressed as
\begin{equation}
\begin{array}{lll}
{\cal L}_{soft} &=& -m_{H^1}^2 H_i^{1\ast}H_i^1-m_{H^2}^2 H_i^{2\ast} H_i^2-
     m_{L^I}^2 \tilde{L}_i^{I\ast} \tilde{L}_i^I-m_{R^I}^2 \tilde{R}^{I\ast}
     \tilde{R}^I-m_{Q^I}^2 \tilde{Q}_i^{I\ast} \tilde{Q}_i^I \\
&& -m_{D^I}^2
     \tilde{D}^{I\ast} \tilde{D}^I-m_{U^I}^2 \tilde{U}^{I\ast} \tilde{U}^I
     + (m_1 \lambda_B \lambda_B+m_2 \lambda_A^i \lambda_A^i+m_3 \lambda_G^a
     \lambda_G^a+h.c.) \\
&& + \{B \mu \epsilon_{ij} H_i^1 H_j^2+B_I \epsilon_I
     \epsilon_{ij} H_i^2 \tilde{L}_j^I+\epsilon_{ij} l_{sI} H_i^1 \tilde{L}_j^I
     \tilde{R}^I \\
&& +d_{sI}(-H_1^1 \tilde{Q}_2^I+C^{IK} H_2^1 \tilde{Q}_1^K)
     \tilde{D}^I+u_{sI}(-C^{KI\ast} H_1^2 \tilde{Q}_2^I + H_2^2 \tilde{Q}_1^I)
     \tilde{U}^I \\
&& +\epsilon_{ij} \lambda_{IJK}^S \tilde{L}_i^I \tilde{L}_j^J
     \tilde{R}^K+\lambda_{IJK}^{S'}(\tilde{L}_i^I \tilde{Q}_2^J \delta^{JK}-
     \tilde{L}_2^I C^{JK} \tilde{Q}_1^J)\tilde{D}^K+\lambda_{IJK}^{S''}
     \tilde{U}^I \tilde{D}^J \tilde{D}^K \\
&& + h.c.\}
\end{array}
\end{equation}
In Eqs.(2.2-4), $L^I, Q^I, H^I$ represents the SU(2) doublets of
lepton, quark and Higgs superfields respectively, while $R^I, U^I,
D^I$ are the singlets of lepton and quark superfields. The
bilinear term $\epsilon_{i j} \epsilon_I H_i^2 \tilde{L}_j^I$ can
give neutrinos and make the diagonalization of mass matrix more
complexity. In the processes considered in this paper, we assumed
 their effects negligible. In this paper we consider only
the lepton number violation, i.e., $\lambda$ and
$\lambda^{\prime}$ are assumed to be non-zero while
$\lambda^{\prime\prime}$ is fixed to be zero.
\par
The main subprocesses for the parent process $pp \rightarrow
 e^+ e^-+X$ are the following three: (1) $u\bar{u}\rightarrow
 e^+ e^-$, (2)$d\bar{d} \rightarrow  e^+ e^-$, (3) $gg \rightarrow
 e^+ e^-$. The Feynman diagrams of subprocesses (1) and (2) contributed
by R-parity conserving MSSM are plotted in Fig.1(a) and Fig.2(a),
respectively. The diagrams of subprocesses (1),(2) involving the
R-parity violating interactions are depicted in Fig.1(b) and
Fig.2($b-c$), respectively. The Fig.3 shows the Feynman diagrams
of subprocess (3) in the R-parity conserving MSSM at the lowest
order. The Feynman diagrams of subprocess (3) involving R-parity
violating interactions are given in Fig.4. For simplicity we do
not give the diagrams which can be obtained by exchanging the
initial gluons in Fig.3 and Fig.4.
\section{Calculation}
\par
Firstly we consider the $ e^+ e^-$ pair production subprocess via
photon and Z boson exchanges in quark-antiquark annihilation as
\begin{equation}
q(p_1)+\bar{q}(p_2) \rightarrow \gamma,Z \to e^-(k_1)+
e^+(k_2),~~(q=u,d).
\end{equation}
The differential Born cross sections at the parton-level for above
subprocesses (1) $u\bar{u}\rightarrow  e^+ e^-$, (2)$d\bar{d}
\rightarrow  e^+ e^-$ in the framework of the R-parity conserving
MSSM, corresponding to the diagrams Fig.1(a) and Fig.2(a)
respectively, are given by
\begin{equation}\label{eq:xsec0}
 d \hat{\sigma}_{MSSM}^{(i)}=  dP_{2f} \frac{1}{12} \sum
 |A^{(i)}_{\gamma}(\hat s,\hat t, \hat u)+ A^{(i)}_Z(\hat s,\hat t, \hat
 u)|^2,~~~(i=1,2)
\end{equation}
where the summation is taken over the spin and color degrees of
freedom of the initial and final states, and $dP_{2f}$ denotes the
two-particle phase space element. The factor $1/12$ results from
the average over the spins and the colors of the incoming partons.
The $A^{(i)}_{\gamma}$ and $A^{(i)}_Z$ represent the amplitudes of
the photon and Z boson exchange diagrams at tree level,
respectively. The Mandelstam kinematical variables in the parton
center of mass system are defined as
\begin{equation}
 \hat{s} = (p_1 + p_2)^2, \quad
  \hat t = (p_1-k_1)^2, \quad
  \hat u = (p_1-k_2)^2,
\label{eq:mandel}
\end{equation}
\par
The expressions of the squared matrix elements for massless
external fermions are
\begin{eqnarray}\label{eq:Zborn}
\sum |{\cal A}^{(i)}_{\gamma}(\hat s,\hat t, \hat u)|^2 & = & 8 \, Q_q^2 \, Q_l^2\, (4\pi
\alpha)^2
\; \frac{(\hat t^2+\hat u^2)}{\hat s^2}~, \nonumber \\[1.mm]
\sum |{\cal A}^{(i)}_Z(\hat s,\hat t, \hat u)|^2 & = & 8 \, \frac{|\chi(\hat s)|^2}{\hat
s^2} \, \left [(v_q^2+a_q^2)(v_l^2+a_l^2) (\hat t^2+\hat u^2) -
4 v_q a_q v_l a_l\, (\hat t^2-\hat u^2) \right ]~, \nonumber \\[1.mm]
2 \sum {\cal R}e ({\cal A}^{(i)}_Z {\cal A}^{(i)*}_{\gamma}) & = & 64 \pi \alpha \, Q_q
\, Q_l \, a_q \, a_l\,  \, \left [ v_q v_l (\hat t^2+\hat u^2)-a_q a_l (\hat t^2-\hat
u^2)\right ] \, \frac{{\cal R}e  \chi(\hat{s})}{\hat s^2}~,
\end{eqnarray}
with
\begin{equation}\label{eq:coup}
v_f=\frac{1}{2 s_w c_w} (I_f^3-2 s_w^2 Q_f), \qquad
a_f=\frac{I_f^3}{2 s_w c_w}\; ,\qquad \chi(\hat s)= \, \frac{4 \pi
\alpha \hat s}{(\hat s-m_Z^2+i\hat s \Gamma_Z /m_Z^2)} \; ,
\end{equation}
where $f=q,l$.
\par In the MSSM with the R-parity lepton number violation, the tree level differential
cross sections for subprocess (1) and (2) can be expressed as
\begin{equation}\label{eq:xsec1}
 d {\hat \sigma_{\rlap/{R}}^{(i)}}(\hat s,\hat t, \hat u)=  dP_{2f} \frac{1}{12} \sum
 |A^{(i)}_{\gamma}(\hat s,\hat t, \hat u)+ A^{(i)}_Z(\hat s,\hat t, \hat u) +
 A_{\rlap/{R}}^{(i)}(\hat s,\hat t, \hat u)|^2~~~
 (i=1,2),\\
\end{equation}
\par
The R-parity violation amplitudes $A_{\rlap/{R}}^{(1)}$ and
$A_{\rlap/{R}}^{(2)}$, which correspond to the subprocess (1)
$u\bar{u}\rightarrow  e^+ e^-$ and (2) $d\bar{d}\rightarrow e^+
e^-$ respectively, can be expressed as:
\begin{equation}
A_{\rlap/{R}}^{(1)}(\hat s,\hat t, \hat u)=A^{(1)}_{\tilde{d}}(\hat s,\hat t, \hat
u),\quad A_{\rlap/{R}}^{(2)}(\hat s,\hat t, \hat u)=A^{(2)}_{\tilde{u}}(\hat s,\hat t,
\hat u)+
A^{(2)}_{\tilde{\nu}}(\hat s,\hat t, \hat u)\\
\end{equation}
where $A^{(1)}_{\tilde{d}}$, $A^{(2)}_{\tilde{u}}$ and
$A^{(2)}_{\tilde{\nu}}$ are just the contributions from the
diagrams Fig.1(b), Fig.2(b) and Fig.2(c), respectively. By using
the relevant Feynmann rules, we can easily write down the
expressions:
\begin{equation}
\begin{array}{lll}
A^{(1)}_{\tilde{d}}(\hat s,\hat t, \hat u)  & = &(\lambda^{\prime}_{11j})^2[\bar{v}
(p_2)Z_{D_j}^{2i}P_R v(k_1)]\frac{i} {\hat{t}-m_{\tilde{d_j}_i}^2}[\bar{u}
(k_2)Z_{D_j}^{2i}P_L u(p_1)] \\
A^{(2)}_{\tilde{u}}(\hat s,\hat t, \hat u)  & = & -(\lambda^{\prime}_{1j1})^2[\bar{u}
(k_1)Z_{U_j}^{1i}P_R u(p_1)]\frac{i} {\hat{u}-m_{\tilde{u_j}_i}^2}[\bar{v}
(p_2)Z_{U_j}^{1i}P_L v(k_2)] \\
A^{(2)}_{\tilde{\nu}}(\hat s,\hat t, \hat u)  & = &-
(\lambda^{\prime}_{j11})(\lambda_{j1 1})[\bar{v} (p_2)P_L
u(p_1)]\frac{i} {\hat{s}-m_{\tilde{\nu_j}}^2+i m_{\tilde{\nu_j}}
\Gamma_{\tilde{\nu_j}} }[\bar{u} (k_1)P_R v(k_2)]\\
&  & - (\lambda^{\prime}_{j11})(\lambda_{j11})[\bar{v} (p_2)P_R
u(p_1)]\frac{i} {\hat{s}-m_{\tilde{\nu_j}}^2+i m_{\tilde{\nu_j}}
\Gamma_{\tilde{\nu_j}}}[\bar{u} (k_1)P_L v(k_2)] \\

\end{array}
\end{equation}
where $Z_{D_k}^{ij}$ and $Z_{U_k}^{ij}$ represent the elements of
the matrices used to diagonalize the down-type squark and up-type
squark mass matrices, respectively.
\par
Now we turn to the calculation of the gluon fusion subprocess. We denote this subprocess
(3) as
\begin{equation}
g(p_1,\alpha,\mu)+g(p_2,\beta,\nu)  \to e^-(k_1)+ e^+(k_2),
\end{equation}
where $\alpha,\beta$ are the color indices of initial gluinos,
respectively. In the MSSM with R-conserving the differential cross
section can be expressed as
\begin{equation}\label{eq:xsec0}
 d {\hat \sigma_{MSSM}}=  dP_{2f} \, \frac{1}{256} \sum
 |A^{(3)}_a(\hat s,\hat t, \hat u)+ A^{(3)}_b(\hat s,\hat t, \hat u)+
 A^{(3)}_c(\hat s,\hat t, \hat u)+
 A^{(3)}_d(\hat s,\hat t, \hat u)+ A^{(3)}_e(\hat s,\hat t, \hat u)|^2  \; ,
\end{equation}
where the sum is again taken over the spin and color degrees of
freedom of the initial gluons and final states, and $dP_{2f}$
denotes the two-particle phase space element. And $A^{(3)}_i$
represent the amplitudes of diagrams in Fig.3(i) (i=a,b,c,d,e). As
we can see from Fig.3, all the diagrams are $Z^{0}$/$\gamma$
exchanging s-channels. Therefore all these contributions can be
neglected. This is the consequence of Furry theorem. The Furry
theorem forbids the production of the spin-one components of the
$Z^{0}$ and $\gamma$, and the contribution from the spin-zero
component of the $Z^{0}$ vector boson coupling with a pair of
leptons is negligibly small. So we have the amplitudes
corresponding to Fig.3(a-e) approaching to zero and the cross
section $\sigma_{MSSM}$ part with the R-parity conserving
interactions (shown in Fig.3) are vanished.
\par
In the case of the R-parity lepton number violation, the
differential cross section of subprocess (3) can be written as
\begin{equation}
\begin{array}{lll}
d {\hat \sigma_{\rlap/{R}}^{(3)}} & = &dP_{2f} \, \frac{1}{256} \sum
 |A^{(3)}_a(\hat s,\hat t, \hat u)+\cdots+ A^{(3)}_e(\hat s,\hat t, \hat u)+
 A^{(4)}_a(\hat s,\hat t, \hat u)+\cdots+ A^{(4)}_i(\hat s,\hat t, \hat u)|^2  \\
& = & dP_{2f} \, \frac{1}{256}\sum
 | A^{(4)}_a(\hat s,\hat t, \hat u)+\cdots+ A^{(4)}_i(\hat s,\hat t, \hat u)|^2  \\
\end{array}
\end{equation}
which has a nonzero result. The total cross section of process $pp
\rightarrow e^+ e^-+X$ with the lower cut value of the $ e^+
e^-$-pair invariant mass $m_{ e^+ e^-}^{cut}$ can be written as:
\begin{equation}
\begin{array}{lll}
\sigma_{ij}(m_{( e^+ e^-)}\geq m^{cut}_{( e^+ e^-)})= \int_{(m^{cut}_{( e^+ e^-)})^2/ s} ^{1} d \tau \frac{d%
{\cal L}_{ij}}{d \tau} \hat{\sigma}_{ij}(\hat{s}=\tau s),\\
\frac{d{\cal L}_{ij}}{d\tau}=\frac{1}{1+\delta_{ij}}\int_{\tau}^{1} \frac{%
dx_1}{x_1} \left\{\left[ f(i,x_1,Q^2)f(j,\frac{\tau}{x_1},Q^2) \right] +\left[
f(j,x_1,Q^2)f(i,\frac{\tau}{x_1},Q^2) \right]\right\}
\end{array}
\end{equation}
where $\sqrt{s}$ and $\sqrt{\hat{s}}$ are the $pp$ collision and
subprocess c.m.s. energies respectively, and $d{\cal L}_{ij}/d
\tau$ is the luminosity of incoming partons. Here, $i,j$ represent
the partons $u,d,\bar{u},\bar{d},g$, $\tau = x_1~x_2$. The
definitions of $x_1$ and $x_2$  can be seen from Ref.\cite{x1x2},
we adopt the CTEQ5 parton distribution function \cite{function}.
In $f(i,x, Q^2)$, factorization scale Q is taken to be
$\sqrt{\hat{s}}$. Then we can obtain the cross section for the
parent process of $pp \rightarrow ij \rightarrow e^+ e^- +X$ with
the dilepton invariant mass $m_{( e^+ e^-)}$ being larger than
$m^{cut}_{( e^+ e^-)}$ :
\begin{equation}
\begin{array}{lll}
\sigma_{ij}(m_{( e^+ e^-)}\geq m^{cut}_{( e^+ e^-)})=
\int_{m^{cut}_{( e^+ e^-)} } ^{\sqrt{s}} d \sqrt{\hat{s}}
 \hat{\sigma}_{ij}(\hat{s})H_{ij}(\hat{s}),\\
H_{ij}(\hat{s})=\frac{1}{1+\delta_{ij}}\int_{\frac{\hat {s}}{s}}^{1} \frac{%
  2dx_1\sqrt{\hat{s}}}{ x_1s} \left\{\left[
f(i,x_1,Q^2)f(j,\frac{\hat{s}}{x_1s},Q^2) \right] +\left[
f(j,x_1,Q^2)f(i,\frac{\hat{s}}{x_1s},Q^2) \right]\right\}
\end{array}
\end{equation}
\par
where $\sigma^{(i)}~(i=1,2,3)$ represent the cross sections of
processes: (1) $pp \rightarrow u\bar{u}\rightarrow  e^+ e^-+X$,
(2) $pp \rightarrow d\bar{d} \rightarrow  e^+ e^-+X$, and (3) $pp
\rightarrow gg \rightarrow  e^+ e^-+X$, respectively. Then the
total cross sections of the $e^+ e^-$ pair production with its
invariant mass larger than $m^{cut}_{( e^+ e^-)}$ in the
R-violating and R-conserving supersymmetric models can be written
as :
\begin{equation}
\begin{array}{lll}
\sigma_{\rlap /{R}}(m_{ e^+ e^-}\geq m^{cut}_{( e^+
e^-)})&=&\sigma^{(1)}_{\rlap /{R}}((m_{( e^+ e^-)}\geq m^{cut}_{(
e^+ e^-)})+\sigma^{(2)}_{\rlap /{R}}(m_{( e^+ e^-)}\geq
m^{cut}_{( e^+ e^-)}) \\
&+& \sigma^{(3)}_{\rlap /{R}}(m_{( e^+ e^-)}\geq m^{cut}_{( e^+ e^-)})\\
\sigma_{MSSM}(m_{( e^+ e^-)}\geq m^{cut}_{( e^+
e^-)})&=&\sigma^{(1)}_{MSSM}(m_{( e^+ e^-)}\geq m^{cut}_{( e^+
e^-)})+\sigma^{(2)}_{MSSM}(m_{( e^+ e^-)}\geq
m^{cut}_{( e^+ e^-)}) \\
&+& \sigma^{(3)}_{MSSM}(m_{( e^+ e^-)}\geq m^{cut}_{( e^+ e^-)})\\
&=& \sigma^{(1)}_{MSSM}(m_{( e^+ e^-)}\geq m^{cut}_{( e^+
e^-)})+\sigma^{(2)}_{MSSM}(m_{( e^+ e^-)}\geq m^{cut}_{( e^+
e^-)})
\end{array}
\end{equation}
where we used $\sigma^{(3)}_{MSSM} \approx 0$, which was mentioned
above. For presentation of the $\rlap/{R}$ effect of the process
$pp\rightarrow  e^+ e^-+X$, we define $\rlap/{R}$ parameter $\eta$
as:
\begin{equation}
\eta= \frac{\sigma_{\rlap/{R}}(m_{( e^+ e^-)}\geq
200GeV)-\sigma_{MSSM}(m_{( e^+ e^-)}\geq
200GeV)}{\sigma_{MSSM}(m_{( e^+ e^-)}\geq 200GeV)}.
\end{equation}

\section{Numerical result and discussion}
\par
 In the numerical calculation we set the input SM parameters to be : $m_u =5MeV$, $m_d
=5MeV$, $m_c =1.2GeV$, $m_s =120MeV$, $m_t =170GeV$, $m_b
=4.2GeV$, $m_Z = 91.187GeV$, $\Gamma_Z = 2.49GeV$. In this work,
we do numerical calculation in the minimal supergravity (mSUGRA)
scenario\cite{msugra}. In this scenario, only five sypersymmetric
parameters  should be given which are named $M_{1/2}$, $M_0$,
$A_0$, sign of $\mu$ and $\tan\beta$, where $M_{1/2}$, $M_0$ and
$A_0$ are the universal gaugino mass, scalar mass at GUT scale and
the trilinear soft breaking parameter in the superpotential
respectively. As we know that the effects of the R-parity
violating couplings on the renormalization group equations(RGE's)
are the crucial ingredient of mSUGRA-type models, and the complete
2-loop RGE's of the superpotential parameters for the
supersymmetric standard model including the full set of R-parity
violating couplings are given in Ref.\cite{Allanach}. But in our
numerical presentation to get the low energy scenario from the
mSUGRA \cite{msugra}, we ignored those effects in the RGE's for
simplicity and use the program ISAJET 7.44. In this program the
RGE's \cite{RGE} are run from the weak scale $m_{Z}$ up to the GUT
scale, taking all thresholds into account and using two loop RGE's
only for the gauge couplings and the one-loop RGE's for the other
supersymmetric parameters. The GUT scale boundary conditions are
imposed and the RGE's are run back to $m_Z$, again taking
threshold into account. The R-parity violating parameters chosen
above satisfy the constraints given by Ref.\cite{B.A}.
\par
We take the mSUGRA input parameters as: $M_{1/2}=150GeV;$, $A_0
=300GeV$, $\tan\beta=4$, $\mu>0$, $m_{t}=170GeV$. The numerical
values of $M_{0}$ will be running in a definite range. The ratio
of $\tilde{\nu}$ decay width  to its mass is taken as
$\Gamma_{\tilde{\nu}}/m_{\tilde{\nu}}=0.07$. Ref.\cite{B.A}
presents the experimental constraints for the coupling parameters
in R-parity violating interactions. According to these upper
limitations, we take the relevant R-violating coupling parameters
$\lambda^{'}_{ijk}$ having the values as
$$
\lambda^{'}_{111}=0.01,
$$
$$
\lambda^{'}_{ijk}=0.04,~~(when~two~of~i,j,k = 1),
$$
$$
\lambda^{'}_{ijk}=0.39,~~(when~two ~or~three~of~i,j,k = 3).
$$
And all of the others $\lambda'_{ijk}$ which we will use in
calculation are chosen to be 0.25. For the related coupling
parameters $\lambda_{ijk}$, we know that the first two indices of
 parameters $\lambda_{ijk}$ are antisymmetric, then the
$\lambda_{ijk}$ should be zero when $i=j$. The others concerned
$\rlap/{R}$ coupling parameters $\lambda_{211}$ and
$\lambda_{311}$ are taken to be -0.18.
\par
The cross section of the subprocess $d \bar d \rightarrow e^+
e^-$, $\hat{\sigma}^{(2)}$ as a function of $\sqrt{\hat{s}}$ is
given in Fig.5. We can see there is a peck around the vicinity of
the $m_{\tilde{\nu}}$ on the curve for the $\rlap/{R}$-MSSM. Since
the energy of this subprocess $\sqrt{\hat{s}}$ from incoming
protons has a continuous spectrum, it can be estimated that the
total cross section and the $\rlap/{R}$ effects of $pp \rightarrow
e^+ e^-+X$ will be greatly enhanced at hadron colliders. The cross
section of subprocess $gg \rightarrow  e^+ e^-$ in
$\rlap/{R}$-MSSM as a function of the c.m.s. energy of colliding
gluons is depicted in Fig.6. Again we can see the resonance
enhancement from $\tilde{\nu}$ exchanging s-channel at the energy
position around $m_{\tilde{\nu}}$. Because of the much larger
luminosity of gluons in the parton distribution function of proton
compared with that of quarks and anti-quarks, the subprocess $gg
\rightarrow  e^+ e^-$ would contribute to the total cross section
of $pp \rightarrow  e^+ e^-+X$ at the LHC to some extent. Fig.7
gives the total cross section of the process $pp \rightarrow  e^+
e^-+X$ as the function of the lower cut value of the $ e^+ e^-$
pair invariant mass $m_{( e^+ e^-)}^{cut}$. From this figure, we
can see that the $\rlap/{R}$ effect on the production rate of $
e^+ e^-$ pair is rather large, and gluon fusion subprocess plays a
significant role. Now we use the parameter $\eta$, which is
defined in Eq.(3.15), to represent the effect of $\rlap/{R}$ of
parent process $pp \rightarrow   e^+ e^- + X$, and we further
denote the $\rlap/{R}$ effect parameters $\eta_{i}~(i=1,2,3)$,
which correspond to the parent processes in which the $\rlap/{R}$
effect is contributed by the Drell-Yan mechanism, gluon fusion and
both subprocesses, respectively. In Fig.8, we depicted the
R-violating effect $\eta$ as a function of universal scalar mass
at GUT scale $M_0$, which is related to the mass of supersymmetric
neutrino $\tilde{\nu}$. From Fig.8 we know that effect of R-parity
lepton number violating on $ e^+ e^-$-pair production at the LHC
is rather large, especially in the $M_0$ range when $M_0<300~GeV$
. The effect parameter of R-parity violation $\eta_3$, which
describes the $\rlap/{R}$ effect for process $ pp \rightarrow  e^+
e^- + X$, may be over the value of 70$\%$. We can see also from
Fig.8 that the most part of the R-parity violating effect is
contributed by the Drell-Yan subprocess, but the part contributed
by gluon fusion subprocess $gg \rightarrow  e^+ e^-$ is also
significant. Fig.9 shows the ratio of the R-parity violating
effects contributed by gluon fusion and Drell-Yan subprocesses, as
a function of universal scalar mass at GUT scale($M_0$). We can
see that the smaller the scalar mass $M_0$ is, the more
contributions to $\rlap/{R}$ effects via subprocess $gg
\rightarrow  e^+ e^-$ are. When $M_0 \sim 200~GeV$, the
contribution to $\rlap/{R}$ effects from gluon fusion can reach
approximately 25$\%$ of that from Drell-Yan subprocess. Fig.10
shows the dependence of the ratio of the $\rlap/R$ effects via
gluon fusion and Drell-Yan mechanism subprocesses on the value of
$\lambda'_{233}(=\lambda'_{333})$. It demonstrates that the ratio
$\eta_2/\eta_1$ is strongly related to the R-parity violating
parameters $\lambda'_{233}(\lambda'_{333})$. The larger
$\lambda'_{233}$ and $\lambda'_{333}$ are, the more important the
contribution to the $\rlap/R$ effect via subprocess $gg
\rightarrow  e^+ e^-$ is. From Fig.8, Fig.9 and Fig.10, we can
conclude that although the gluon fusion subprocess $gg \rightarrow
e^+ e^-$ has no tree-level Feynman diagram, its contribution to
$\rlap/{R}$ effect is also significant comparing to that via
Drell-Yan subprocess in some parameter space of the MSSM with
R-parity violating. The reason are in two fields: Firstly, it is
due to the large gluon luminosity in hadron collider. Secondly, in
the loop-order Feynman diagrams for gluon fusion subprocess, there
exist the R-parity lepton number violation interactions involving
the third generation particles. While the experimental upper
limitations of these coupling strengths\cite{B.A} are much larger
than those, their interacting particles involving only the members
of the first two generations.
\par
\section{Summary}

 We studied the effects of the R-parity lepton number
violation in the minimal supersymmetric standard model (MSSM) on
the parent process $pp \rightarrow   e^+ e^- + X$ at the CERN
Large Hadron Collider (LHC). The contribution from the one-loop
induced gluon fusion subprocess $gg \rightarrow  e^+ e^-$, may
contribute to R-parity violating effect significantly in comparing
with that via Drell-Yan subprocess. So in studying the effect of
R-parity lepton number violation on $ e^+ e^-$-pair production
process, we should figure out the contributions both from
Drell-Yan subprocess $q \bar q \rightarrow e^+ e^-$ and gluon
fusion subprocess $gg \rightarrow e^+ e^-$. Our calculation shows
that the R-parity lepton number violating effect from gluon fusion
subprocess can reaches 30$\%$ of the corresponding part via
Drell-Yan subprocess. We find also that the dependences of
R-violating effect on the sneutrino mass and the coupling strength
of the R-parity violating interactions are obvious.

\vskip 4mm \noindent{\large {\bf Acknowledgement:}}
\par
This work was supported in part by the National Natural Science Foundation of China, and
a grant from the University of Science and Technology of China.

\vskip 10mm
\begin{flushleft} {\bf Figure Captions} \end{flushleft}

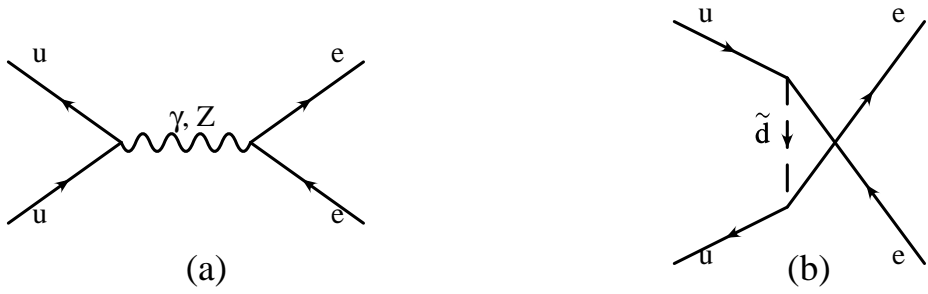
{\bf Fig.1} The relevant Feynman diagrams for the subprocess $u \bar{u} \rightarrow e^+
e^-$ in the MSSM at the tree-level: (a) for the Feynman diagrams for R-parity conserved
MSSM part; (b) for the Feynman diagrams for R-parity violation MSSM part.
\par
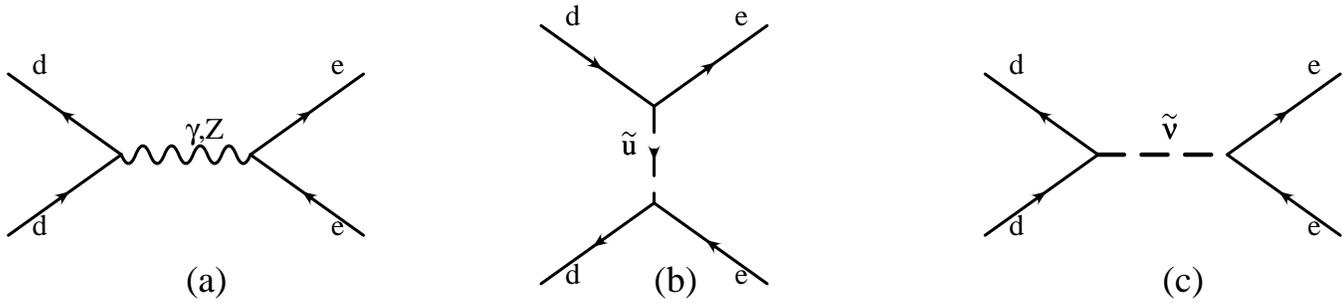
{\bf Fig.2} The relevant Feynman diagrams for the subprocess $d \bar{d} \rightarrow e^+
e^-$ in the MSSM at the tree-level: (a) for the Feynman diagrams for R-parity conserved
MSSM part; (b) and (c) are for the Feynman diagrams for R-parity violation MSSM part.
\par
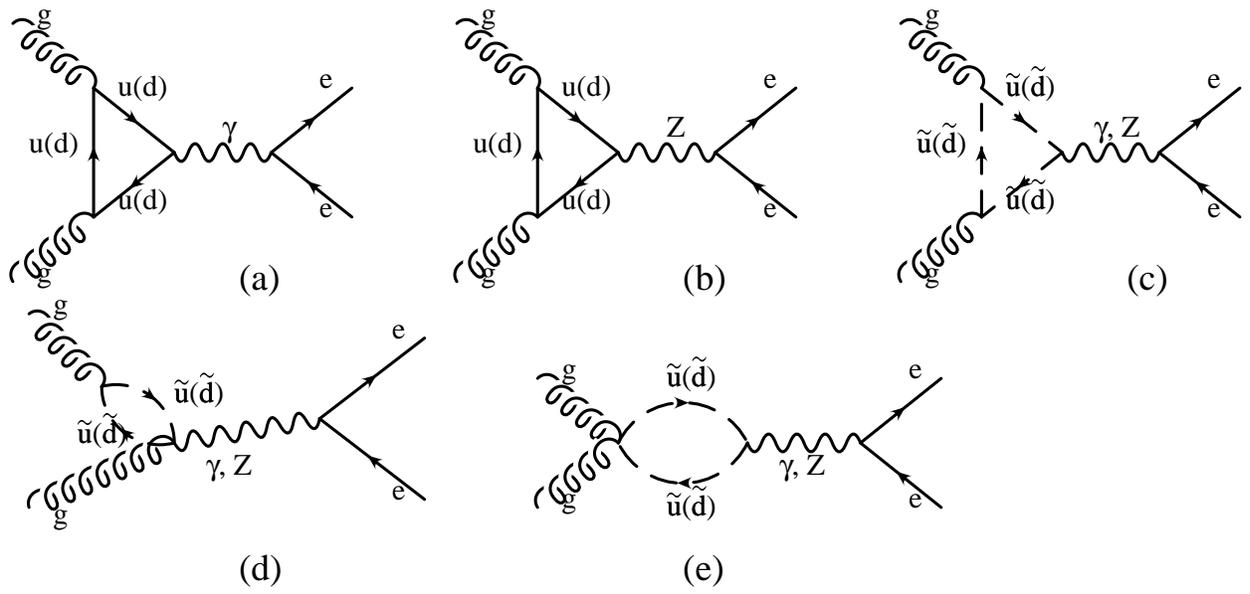
{\bf Fig.3} The relevant Feynman diagrams for the subprocess $gg
\rightarrow   e^+ e^-$ for the R-parity conserved MSSM part in the
MSSM at the lowest level.
\par
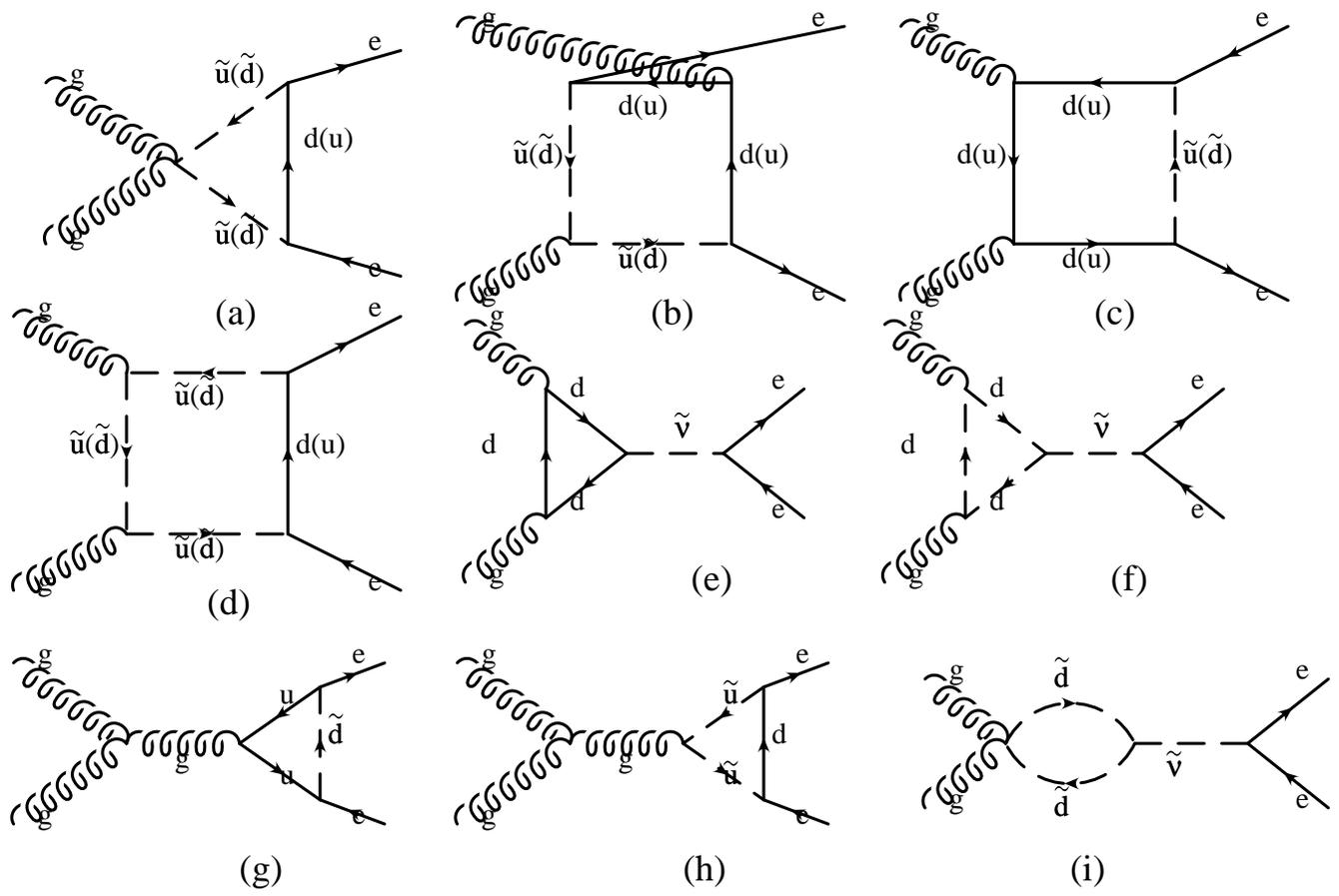
{\bf Fig.4} The relevant Feynman diagrams for the subprocess $gg
\rightarrow   e^+ e^-$ for the R-parity violation MSSM part in the
MSSM at the lowest level.
\par
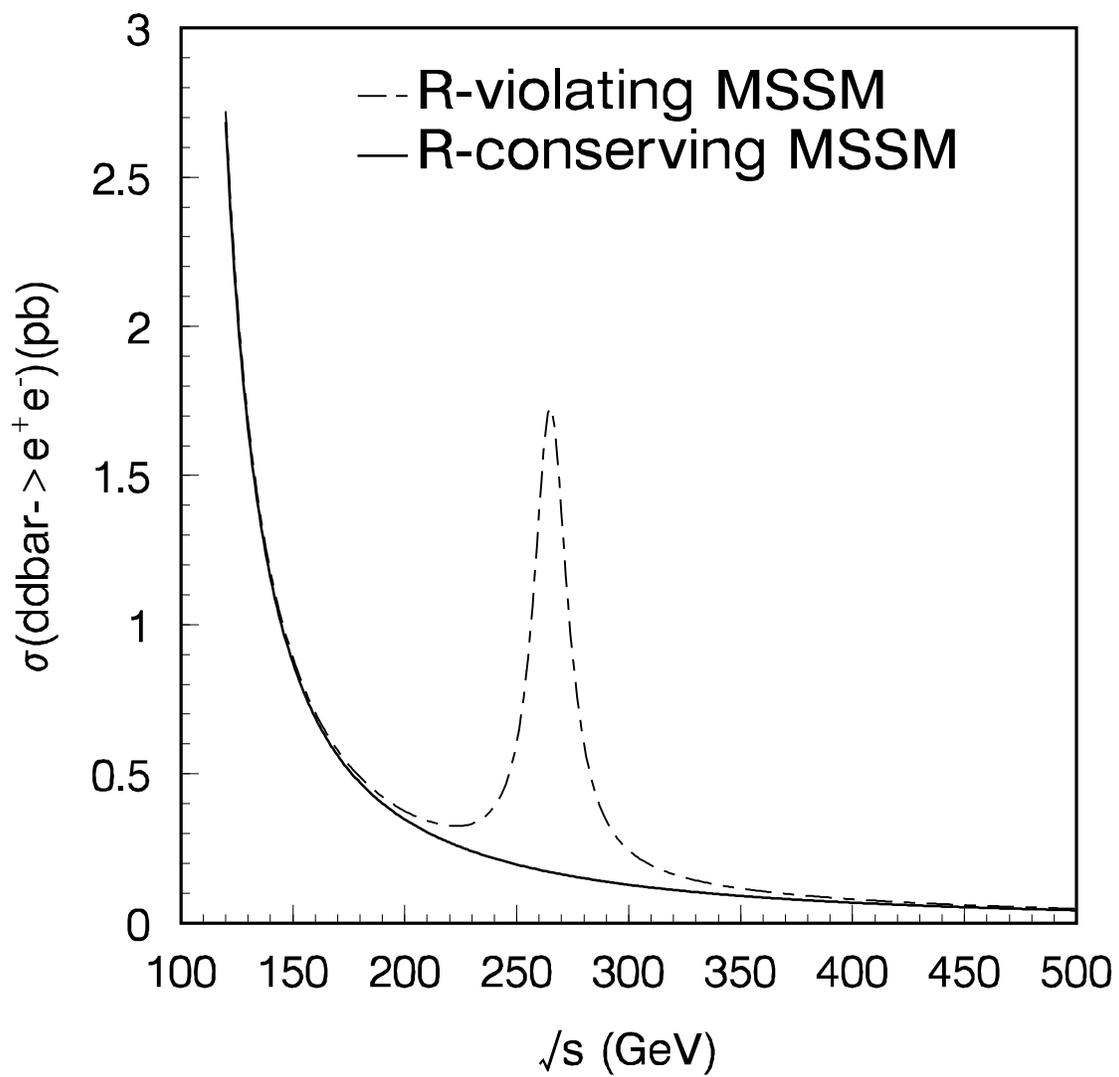
{\bf Fig.5} The cross section of subprocess $d\bar d\rightarrow
 e^+ e^-$ at the LHC as the function of $\sqrt{\hat{s}}$, with the
colliding proton-proton energy $\sqrt{s}=14TeV$ and
$M_{0}=250~GeV$. The full line is for R-conserving MSSM, the
dashed line for R-violating MSSM.
\par
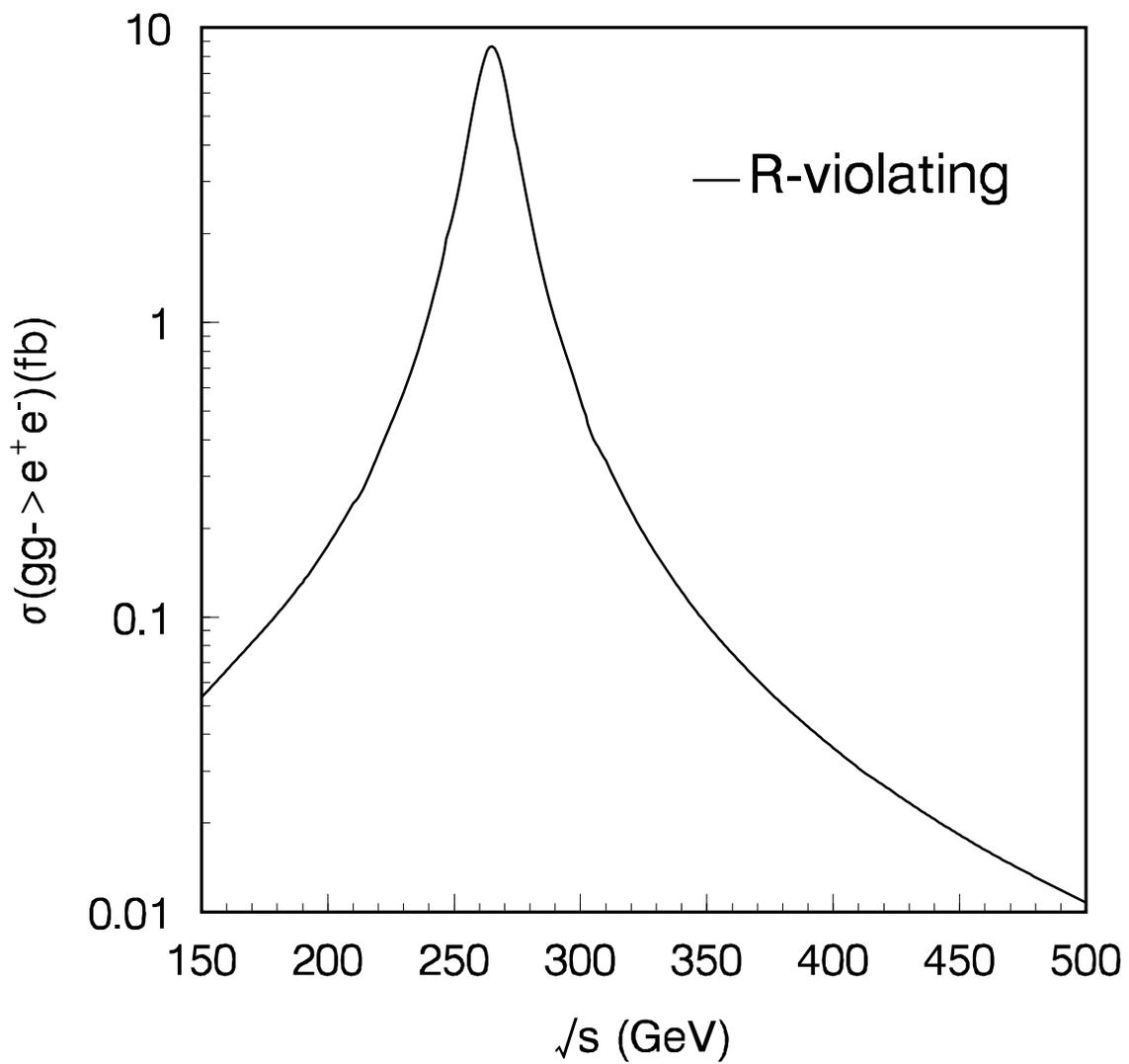
{\bf Fig.6} The cross section of subprocess $gg\rightarrow  e^-
e^+$ at the LHC as the function of $\sqrt{\hat{s}}$, with the
colliding energy $\sqrt{s}=14TeV$ and $M_{0}=250GeV$.
\par
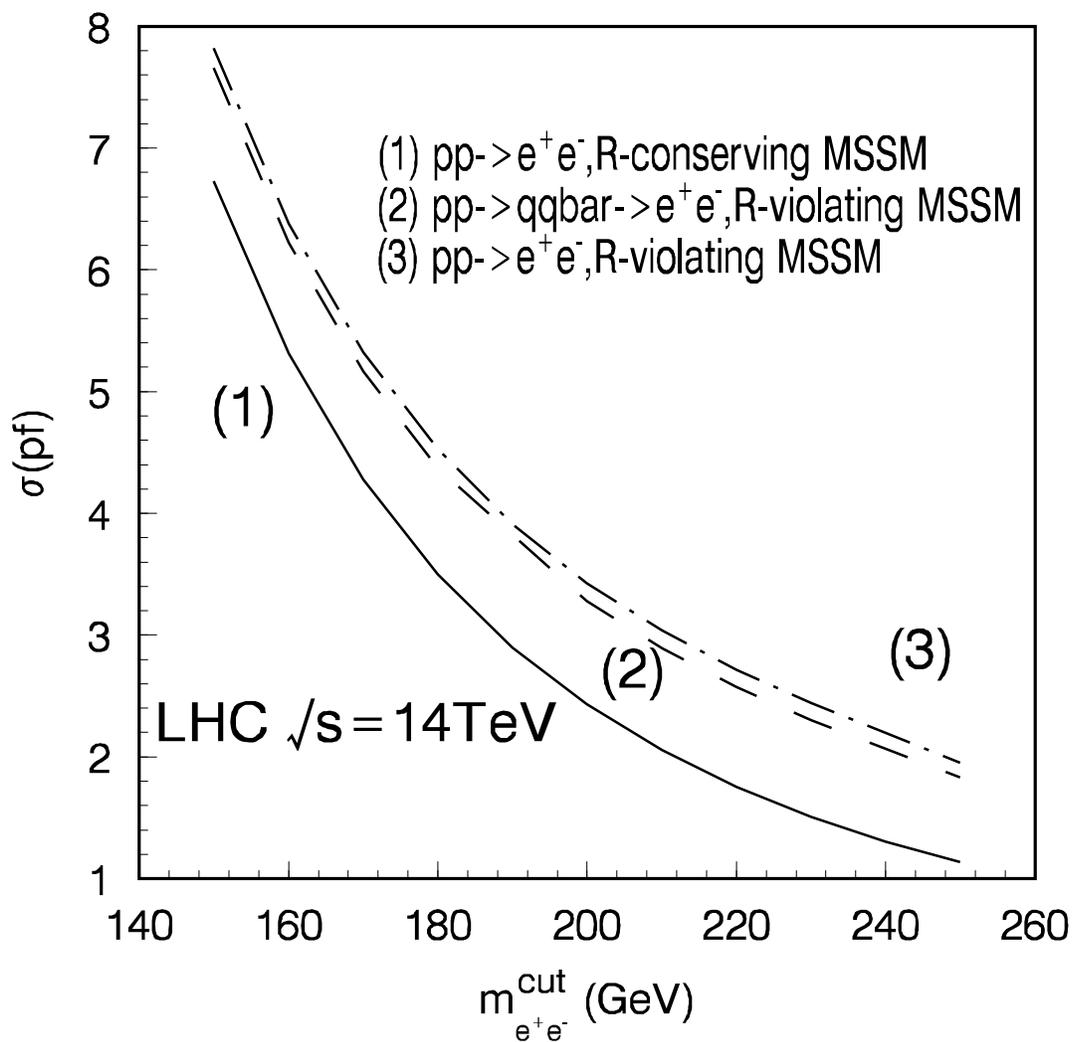
{\bf Fig.7} The cross section of process $pp\rightarrow   e^+
e^-+X$ at the LHC as the function of $m_{ e^+ e^-}^{cut}$, with
the colliding energy $\sqrt{s}=14TeV$ and $M_{0}=250GeV$. The
full-line is for the R-conserving MSSM. The dashed-line is for the
R-violating MSSM, where only $\rlap/{R}$ effect via $q\bar
q\rightarrow
 e^+ e^-$ is taken into account. The dotted-dashed-line is for the
R-violating MSSM, where the effects via both Drell-Yan and gluon
fusion subprocesses are taken into account.
\par
{\bf fig.9} R-parity violating effect parameters
$\eta_{i}(i=1,2,3)$ as the functions of $M_{0}$ . The full-line is
for $\eta_{1} $ (for case 1: only $\rlap/{R}$ effect via $q
\bar{q} \rightarrow  e^+ e^-$ is taken into account). The
dotted-dashed-line for $\eta_{2}$ (for case 2: only $\rlap/{R}$
effect via $q\bar q\rightarrow  e^+ e^-$ is taken into account).
The dashed-line for $\eta_{3}$ (for case 3: the effects via both
Drell-Yan and gluon fusion subprocesses are taken into account).
\par
{\bf Fig.9} The ratio ($\eta_{2}/\eta_{1}$) of the R-violating
parameters as a function of $M_{0}$ , which shows the importance
of the contributions from subprocesses $q\bar q\rightarrow  e^+
e^-$ and $gg \rightarrow  e^+ e^-$.
\par
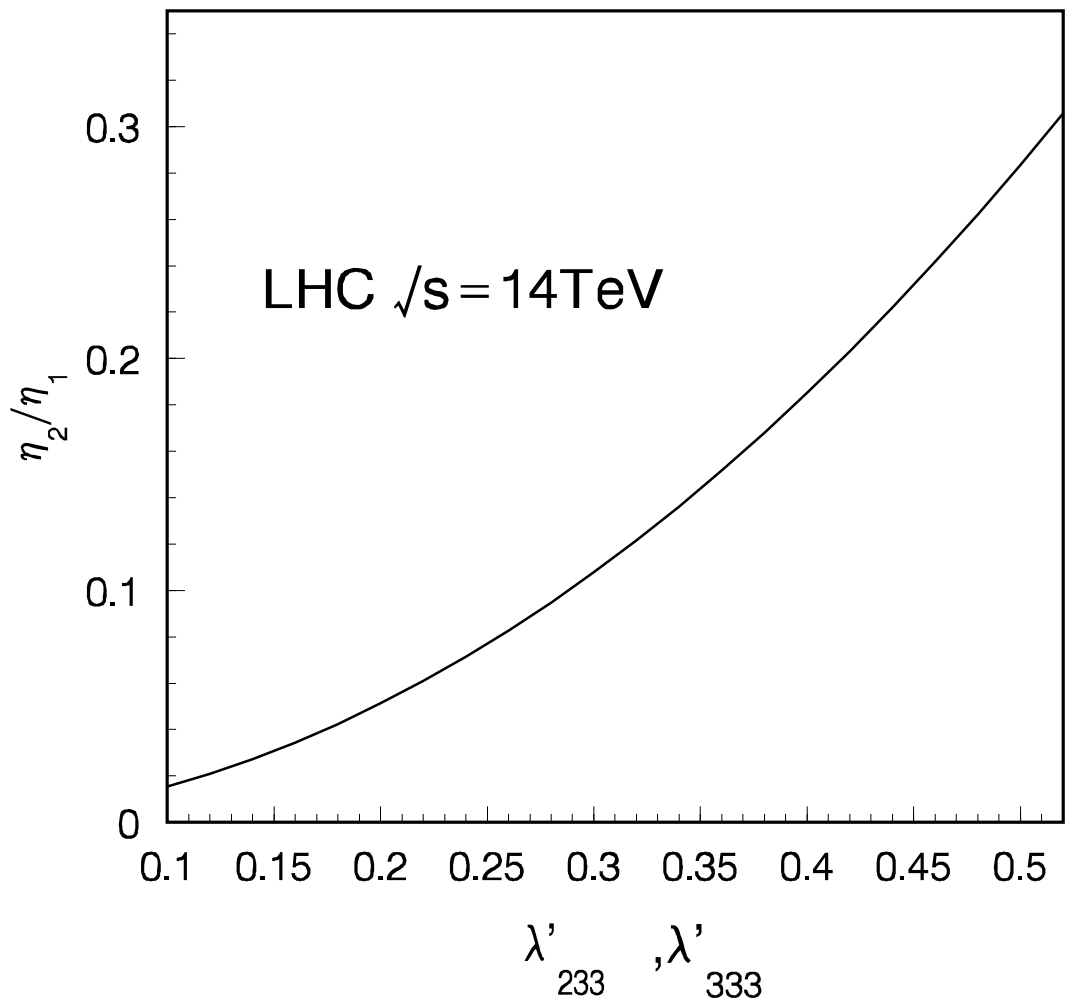
{\bf Fig.10} The ratio ($\eta_{2}/\eta_{1}$) as a function of the R-violating parameters
$\lambda'_{233}(\lambda'_{333})$(where we take $\lambda'_{233}=\lambda'_{333}$) with
$M_0=250GeV$.
\par


\begin{thebibliography}{s25}
\bibitem{beyondSM}  H. E. Haber and G. L. Kane, Phys. Rep. {\bf 117} (1985)
75.
\bibitem{rdef}  G. Farrar and P. Fayet, Phys. Lett. {\bf B76} (1978) 575;
Phys. Lett. {\bf B79} (1978) 442 L. Ibanez and G.G. Ross, Phys. Lett. {\bf %
B260} (1991) 291, Nucl. Phys. {\bf B368} (1992) 3; J. Ellis, S. Lola and G.G. Ross, Nucl.
Phys. {\bf \ B526} (1998) 115.
\bibitem{U.B} U.Baur, O.Brein, W.Hollick, C.Schappacher, D.Wacheroth,
Phys.Rev. D65 (2002) 033007.
\bibitem{F.A1} F.Abe $et$ $al.$(CDF Collaboration), Phys. Rev. D49, R1(1994).
\bibitem{F.A2} F.Abe $et$ $al.$(CDF Collaboration), Phys. Rev. D59, 052002(1999).
\bibitem{F.A3} F.Abe $et$ $al.$(CDF Collaboration), Phys. Lett. 75, 11(1995)
and Phys. Rev.D52, 4784(1995), T affolder $et$ $al.$ (CDF Collaboration), Phys. Rev.
D64,052001(2001).
\bibitem{S.A} S.Abachi $et$ $al$. (D0 Collaboration), Phys, Rev,
Lett. 77,3309(1996), B. Abbott $et$ $al$. (D0 Collaboration), Phys, Rev, D58,
12002(1998), Phys. Rev. D58, 092003(1998); Phys, Rev, Lett 80,3008(1998); Phys. Rev.
Lett. 84, 222(2000); Phys, Rev, D62, 092006(2000).
\bibitem{R.B} R. Brock $et$ $al$. hep-ex/001009(November 2000), in
proceeding of  the workshop on QCD and Weak Boson Physics in $Run II$, Fermilab,1999,
FERMILAB-Pub-00/297(2000), eds, U. Baur, R.K. Ellis and D.Zeppenfeld, p78.
\bibitem{C.A} C.Albajar et al. (UA1 Collaboration).Z. Phys
C44,15(1989); F. Abe et al.(CDF Collaboration), Phys. Rev. Lett. 67, 1502(1991); P.
Hur(CDF Collaboration); Ph. D. Thesis, University of Illinois at Urbana-Champaign, 1990;
T. Affolder $et$ $al$. (CDF Collaboration), Phys.Rev.Lett. 87 (2001) 131802,
hep-ex/0106047.
\bibitem{S.A2} S,Abachi $et$ $al$.(D0 Collaboration), Phys. Rev.
Lett. 75, 1456 (1995); B,Abbott $et$ $al$.(D0 Collaboration), Phys. Rev. D61,
072001(2000).
\bibitem{F.A4} F,Abe $et$ $al$.(CDF Collaboration), Phys, Rev, Lett, 220(1994)
and Phys, Rev, D52, 2624(1995).

\bibitem{F.A5}G. Giudice, R. Rattazzi, and J. Wells, Nucl. Phys.
B544,3(1999); T. Han, J.D. Lykken and R.-J. Zhang, Phys. Rev. D59,105006(1999); J.L.
Hewett, Phys. Rev. Lett. 82, 4765(1999).

\bibitem{F.A7}B. Abbot $et$ $al$.(D0 Collaboration), Phys. Rev. Lett. 86,1156(2001).
\bibitem{F.A6}F. Abe  $et$ $al$.(CDF Collaboration), Phys. Rev
Lett. 79, 2191 (1997); Phys. Rev. D51, 949 (1995); Phys. Rev. Lett. 68, 1463 (1992); S.
Abachi $et$ $al$.(D0 Collaboration), Phys. Lett. B385, 471(1996); V.M. Abazov $et$
$al$.(D0 Collaboration), Phys. Rev. Lett. 87, 061802 (2001); A. Bedek and U. Baur,
Eur.Phys.J. C21 (2001) 607, hep-ph/0102160.
\bibitem{chou}D. Choudhury and R.M. Godbole, "Probing
$\rlap/R_{p}$coupling through indirect effects on Drell-Yan production at the LHC", on
"The SUSY working Group: Summary Report", p65, hep-ph/0005142.
\bibitem{F.A8}F. Abe et al. (CDF Collaboration), Phys. Rev. lett.
79, 2198(1997); B. Abbott, et al. (D0 collaboration), Phys. Rev. Lett. 82, 4769(1999).
\bibitem{B.A}    B. Allanach, H. Baer, S. Banerjee, E.L. Berger et al.,
                 hep-ph/9906224, and the references herein; B.C. Allanch,
                 A. Dedes, and H.K. Dreiner, Phys. Rev. {\bf D60},(1999)
                 075014.
\bibitem{Yin}  P. Roy, TIFR/TH/97-60; D.K. Ghosh, S. Raychaudburi and
K. Sridbar, Phys. Lett. B396(1997)177.;X. Yin, W.G. Ma, L.H. Wan, L. Han and Y. Jiang
Comm. Theor. Phys. 36 (2001) 553.
\bibitem{x1x2}  Y. Jiang, W.G. Ma, L. Han, Z.H. Yu and H. Pietschmann, Phys.
Rev. {\bf D62} (2000)035006.
\bibitem{function}  H.L.Lai, J.Huston, and S.Kuhlmann , Eur.Phys.J. C12 (2000)375,
hep-ph/9903282.
\bibitem{msugra} M. Drees and S. P. Martin, MAD-PH-879, UM-TH-95-02,hep-ph/9504324.
\bibitem{Allanach} B.C. Allanach,A. Dedes and H.K. Dreiner, Phys. Rev. D60 (1999)
            075014, hep-ph/9906209.
\bibitem{RGE} V. Barger, M. S. Berger and P. Ohmann, Phys. Rev. {\bf D47},
              1093(1993), {\bf D47}, 2038(1993); V. Barger, M. S. Berger,
              P. Ohmann and R. J. N. Phillips, Phys. Lett. {\bf B314},
              351(1993); V. Barger, M. S. berger and P. Ohmann, Phys. Rev.
              {\bf D49}, 4908(1994).

\end{thebibliography}
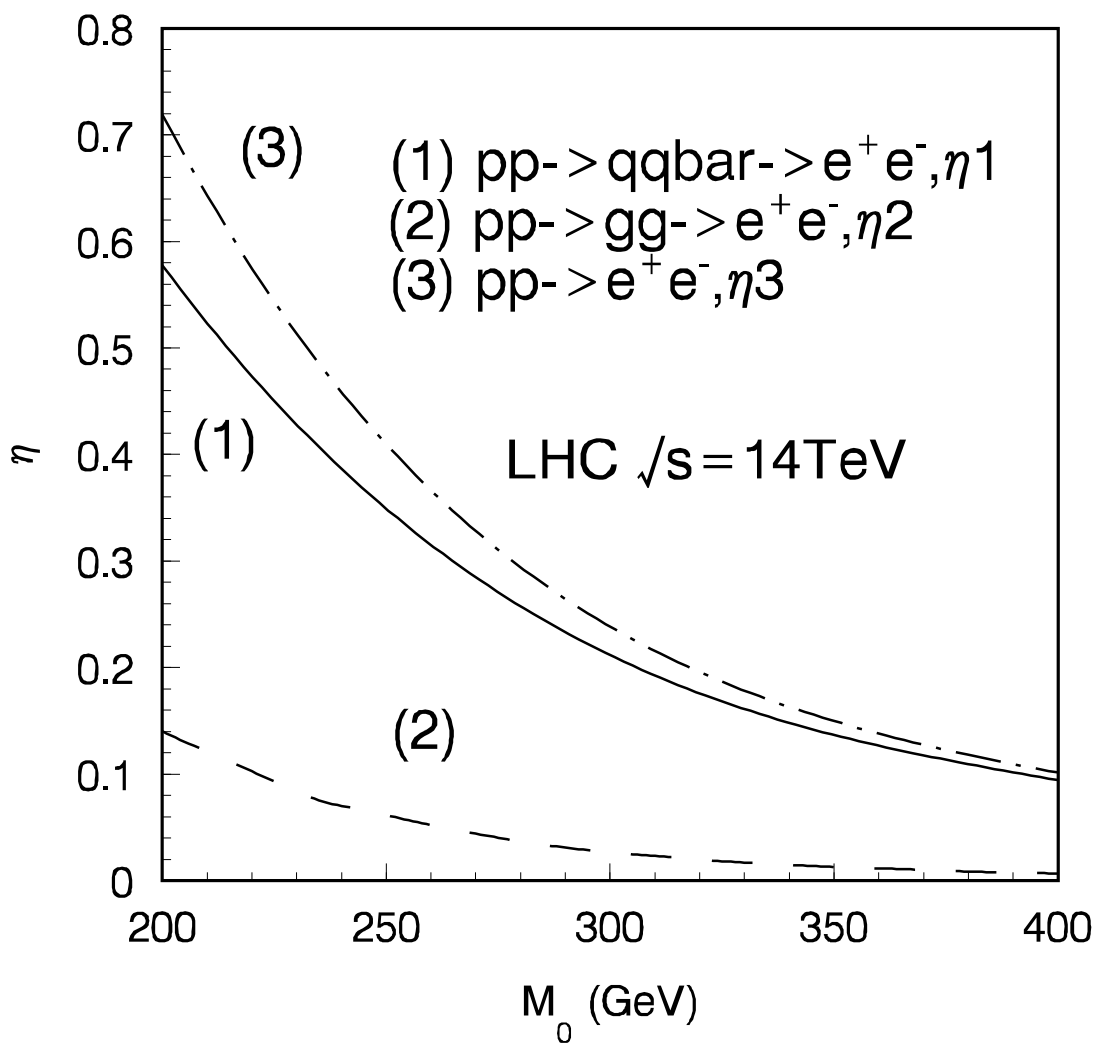
\end{document}